\documentclass[11pt,a4paper]{article}
\usepackage{amssymb}
\usepackage{amsmath}
\usepackage{graphicx}
\usepackage{subfig}
\usepackage[english]{babel}
\newcommand{\be}{\begin{equation}}
\newcommand{\ee}{\end{equation}}
\newcommand{\nn}{\nonumber}

\newcommand{\I}{{\cal I}}
\newcommand{\F}{{\cal F}}

\newcommand{\N}{{\cal N}}

\renewcommand{\O}{{\cal O}}

\newcommand{\rmd}{{\rm d}}
\newcommand{\ts}{\textstyle}

\newcommand{\half}{{\tfrac{1}{2}}}
\newcommand{\quar}{{\tfrac{1}{4}}}
\newcommand{\pr}{\partial}

\newcommand{\vphi}{\varphi}

\newcommand \bz{{\bar z}}
\newcommand \bh{{\bar h}}

\newcommand \ollr{{\raise 8pt\hbox{$\leftrightarrow  \! \! \! \! \! \!$}}}

\csname @addtoreset\endcsname{equation}{section}
\begin{document}

\begin{titlepage}
\thispagestyle{empty}
\begin{flushright}
\small
DAMTP/12-37\\
arXiv:1205.1941[hep-th]\\
\today \\
\normalsize
\end{flushright}
\vskip 3cm
\centerline{\LARGE{\bf Conformal Blocks for Arbitrary Spins in Two Dimensions}}
\bigskip
\vskip 3cm
\centerline{H. Osborn\footnote{ho@damtp.cam.ac.uk}}
\vskip 1cm
\centerline{Department of Applied Mathematics and Theoretical Physics,}
\centerline{Wilberforce Road, Cambridge, CB3 0WA, England}

\date{latest update: \today}
\begin{abstract}
Conformal blocks for the finite dimension conformal group $SO(2,2)$ 
for four point functions for fields with arbitrary spins
in two dimensions are obtained by evaluating an appropriate integral. The
results are just products of hypergeometric functions of the conformally
invariant cross ratios formed from the four complex coordinates. Results
for scalars previously obtained are a special case. Applications to four point
functions involving the energy momentum tensor are discussed.
\end{abstract}

\thispagestyle{empty}
\end{titlepage}
\setcounter{footnote}{0}
\newpage

There has been a resurgence of interest in applying bootstrap methods
to understanding the structure of conformal and superconformal field theories.
Bounds on the anomalous dimensions depending on general
principles such as conformal symmetry the operator product expansion
and crossing have been analysed. Most  applications have been to
four dimensional theories \cite{Rychk, Poland} 
but very recently \cite{Ising} the three dimensional Ising model
has been investigated with remarkable indications that this may become a
very effective method for determining critical exponents.

In order to apply the bootstrap equations it is necessary to be able to
determine the conformal blocks representing  the contributions of particular
conformal primary operators and their descendants to a four point function for
conformal primary operators when the operator product expansion is applied 
to distinct pairs of operators. Equating the presumed convergent sums over 
conformal blocks in different channels at convenient points 
leads, together with positivity conditions arising from unitarity, to
non trivial constraints. In general conformal blocks depend on two conformal
invariants $u,v$ and are the extensions to the non compact conformal group
$SO(d,2)$, or $SO(d+1,1)$,
of two variable harmonic polynomials for the corresponding compact $SO(d+2)$.
In two and four dimensions simple expressions
for the conformal blocks for four point functions for scalar operators were
found in terms of hypergeometric functions of variables $z,\bz$ which are
simply defined in terms of $u,v$ \cite{DO}. Recently these results have
been extended to include cases when the four point function  involves
fields with spin \cite{Joao, paulos, Duffin} although concise
simple generalisations of the spinless case for arbitrary spins have not so 
far been found.

In this note we show how conformal blocks for four point functions for
arbitrary spins may be found in two dimensions. The results are just a
simple extension of the spinless case. Of course two dimensions is very
special, not least because the associated rotational group $SO(2)$ has
only one dimensional representations labelled by spin or helicity $s$. The
difficulties of handling multi-index tensors or spinors are part of the
complexities that arise in the four dimensional case but progress has been
made in handling these  \cite{Joao, Duffin} using twistor methods.

Our starting point is an integral representation which immediately gives
rise to a conformal block. The conformally covariant integrals which are 
relevant  in two dimensions have the form, written in terms
of standard complex variables $z,\bz$ with $\rmd^2 z = \rmd{\it Re}z ,
\rmd{\it Im}z$, 
\be
I_n = \frac{1}{\pi} \int \rmd^2 z \;\prod_{i=1}^n 
\frac{1}{(z-z_i)^{\, \alpha_i}} \, 
\frac{1}{(\bz- \bz_i)^{\, {\bar \alpha}_i}}  \, ,  \quad
\sum_{i=1}^n \alpha_i = {\sum_{i=1}^n} {\bar \alpha}_i = 2 \, , \ \
\alpha_i - {\bar \alpha}_i \in {\mathbb Z} \, .
\label{In}
\ee
This was evaluated in general in \cite{recur} for $n=2,3,4$. 

In two dimensions the conformal primary fields $\vphi(z,\bz)$ are labelled
by $h,\bh$ with the scale dimension $\Delta = h+\bh$ and spin 
$s = h - \bh$ \cite{cftbook} and we require $2s \in {\mathbb Z}$,
with $\vphi$ bosonic/fermionic according to whether $2s$ is even/odd.
For any $\vphi(z,\bz)$ conjugation gives a conformal primary field
 ${\bar \vphi}(z,\bz)$ with
$h \leftrightarrow \bh$. For real scalars $h=\bh$ we may impose
$\vphi(z,\bz) = \vphi(\bz,z)$. Corresponding to $\vphi(z,\bz)$ we define
a dual or shadow field
\be
{\tilde \vphi}(z,\bz) = K_{h,\bh}\,  \frac{1}{\pi} \int \rmd^2 y \;
\frac{1}{(z-y)^{2-2h} ( \bz - {\bar y})^{2-2\bh}} \, \vphi(y,{\bar y}) \, ,
\ee
which is a conformal primary with ${\tilde h} = 1-h, {\tilde \bh}=1-\bh$. 
Choosing
\be
K_{h,\bh} = \frac{\Gamma(2-2\bh)}{\Gamma(2h-1)} = (-1)^{2(h-\bh)} \, 
\frac{\Gamma(2-2 h)}{\Gamma(2\bh-1)} \, ,
\ee
ensures, using \eqref{In} for $n=2$, that 
${\tilde{\tilde \vphi}} =  (-1)^{2(h-\bh)} \vphi$.

The operator product expansion for $\vphi_1 \vphi_2$ contains all 
operators $\O$ for which there is a non zero three point function. 
For $\vphi_1,\vphi_2, \O$ conformal primaries\footnote{Our use  of 
the term conformal primary, requires just
$L_1 \vphi_i(0,0) = {\bar L}_1 \vphi_i(0,0) = 0$ as well as
$L_0 \vphi_i(0,0) = h_i \vphi_i(0,0)$, ${\bar L}_0 \vphi_i(0,0) = \bh_i \vphi_i(0,0)$
where $L_\pm,L_0$ and ${\bar L}_\pm, {\bar L}_0$, with algebra 
$\mathfrak{sl}_2 \times \mathfrak{sl}_2$, are the conformal generators. This use
of the notion of conformal primary agrees with that in higher dimensions when
it is necessary that $\vphi(0)$ is annihilated by the conformal generator $K_a$.
In two dimensional conformal field theories what is commonly referred to as a conformal 
primary field  $\vphi$, where $L_n \vphi_i(0,0) = {\bar L}_n \vphi_i(0,0) = 0$
for all $n>0$, is here called a Virasoro primary.}
this is determined by conformal symmetry so that
\begin{align}
\big \langle \vphi_1(z_1,\bz_1) \,\vphi_2(z_2,\bz_2)\, \O( z, \bz) \big \rangle
= {}& (-1)^{2(h-\bh)} \big \langle \O( z, \bz)\, \vphi_1(z_1,\bz_1) \,
\vphi_2(z_2,\bz_2) \big \rangle \nn \\
= {}& C_{12\O} \, \F_{12}^{h\,\bh}(z,\bz) \, ,
\label{ppO}
\end{align}
where we require $h_1-\bh_1 + h_2-\bh_2 + h-\bh \in {\mathbb Z}$ and
\begin{align}
 \F_{12}^{h\, \bh}(z,\bz) = {}& \frac{1}{z_{12}{\!}^{h_1+h_2-h}\, 
(z_1-z)^{h+h_{12}} (z_2-z)^{h-h_{12}}}\,
 \frac{1}{\bz_{12}{\!}^{\bh_1+\bh_2-\bh}\, (\bz_1-\bz)^{\bh+\bh_{12}}
 (\bz_2-\bz)^{\bh-\bh_{12}}} \nn \\
={}& (-1)^{2(h-\bh)} \,  \frac{1}{z_{12}{\!}^{h_1+h_2-h}\, (z- z_1)^{h+h_{12}}
 (z-z_2)^{h-h_{12}}}\nn \\
\noalign{\vskip -8pt}
&\hskip 5cm {}\times  \frac{1}{\bz_{12}{\!}^{\bh_1+\bh_2-\bh}\, 
(\bz-\bz_1)^{\bh+\bh_{12}} (\bz-\bz_2)^{\bh-\bh_{12}}} \, ,
\label{threep}
\end{align}
defining 
$z_{12}=z_1-z_2, \, h_{12}=h_1-h_2$ and similarly for $\bz_{12},\bh_{12}$.
Using the result for $I_3$ as defined in \eqref{In}
\begin{align}
 K_{h,\bh}\,  \frac{1}{\pi} \int \rmd^2 y \;
\frac{1}{(z-y)^{2-2h} ( \bz - {\bar y})^{2-2\bh}} \,& \F_{12}^{h\, \bh}
(y,{\bar y}) \nn \\
\noalign{\vskip -4pt}
={}& \frac{\Gamma(1-h-h_{12})\, \Gamma(1-\bh+\bh_{12})}{\Gamma(\bh+\bh_{12})\, 
\Gamma(h- h_{12})} \,  \F_{12}^{1-h\,1-\bh}(z,\bz) \, .
\end{align}

For  the four point function
$\big \langle \vphi_1(z_1,\bz_1) \,\vphi_2(z_2,\bz_2)  \, \vphi_3(z_3,\bz_3) \,
\vphi_4(z_4,\bz_4)\big \rangle $, requiring the spins to be constrained by
  $\sum_{i=1}^4 (h_i-\bh_i) \in {\mathbb Z}$, the conformal block
corresponding  to the operator product expansion of $\vphi_1\vphi_2$ and 
$\vphi_3\vphi_4$ both containing the operator $\O$ may be
determined by evaluating the conformally covariant integral 
\begin{align}
& \frac{\Gamma(\bh+\bh_{12})\, \Gamma(1-\bh+\bh_{34})}{\Gamma(1-h-h_{12})\,
\Gamma(h- h_{34})} \, \frac{1}{\pi} \int \rmd^2 z \; 
 \F_{12}^{h\, \bh}(z,\bz) \;  \F_{34}^{1-h\, 1-\bh}(z,\bz) \nn \\
&{}=  (-1)^{2(h-\bh)} \,  \frac{\Gamma(1-\bh+\bh_{12})\, 
\Gamma(\bh+\bh_{34})}{\Gamma(h - h_{12})\,
\Gamma(1- h- h_{34})} \, \frac{1}{\pi} \int \rmd^2 z \; 
\F_{12}^{1-h\, 1-\bh}(z,\bz) \;  \F_{34}^{h\, \bh}(z,\bz) \nn \\
&{}= \frac{1}{z_{12}{\!}^{h_1+h_2}\, z_{34}{\!}^{h_3+h_4}} \, 
\Big ( \frac{z_{24}}{z_{14}} \Big )^{h_{12}} 
\Big ( \frac{z_{14}}{z_{13}} \Big )^{h_{34}}\,
 \frac{1}{\bz_{12}{\!}^{\bh_1+\bh_2}\,  \bz_{34}{\!}^{\bh_3+\bh_4}}
\Big ( \frac{\bz_{24}}{\bz_{14}} \Big )^{\bh_{12}} 
\Big ( \frac{\bz_{14}}{\bz_{13}} 
\Big )^{\bh_{34}}\, \I(\eta,{\bar \eta}) \, ,
\label{res1}
\end{align}
for $\eta, {\bar \eta}$ conformal invariants given here by
\be
\eta = \frac{z_{12} \, z_{34}}{z_{13} \, z_{24}}  \, , \qquad
{\bar \eta} = \frac{\bz_{12} \, \bz_{34}}{\bz_{13} \, \bz_{24}} \, .
\ee
The evaluation of $I_4$ given in \cite{recur} leads to
\begin{align}
\I(\eta,{\bar \eta}) = {}& \frac{(-1)^{2(h-\bh)}}{K_{1-h,1-\bh}} \, 
\frac{\Gamma(\bh+\bh_{12})\,
\Gamma(\bh+\bh_{34})}{\Gamma(1- h - h_{12})\, \Gamma(1- h- h_{34})} \,
F_{h\, \bh}(\eta,{\bar \eta})\nn \\
&{} +  \frac{1}{K_{h,\bh}} \, \frac{\Gamma(1-\bh+\bh_{12})\,
\Gamma(1-\bh+\bh_{34})}{\Gamma(h - h_{12})\, \Gamma(h- h_{34})} \,
F_{1-h\, 1-\bh}(\eta,{\bar \eta}) \, ,
\label{res2}
\end{align}
with
\be
F_{h\, \bh}(\eta,{\bar \eta})= \eta^h F (h-h_{12},h+h_{34};2h; \eta ) \;
{\bar \eta}{}^\bh F (\bh-\bh_{12},\bh+\bh_{34};2\bh; {\bar \eta} ) \, .
\label{block}
\ee
The symmetry of  \eqref{res2} under $h\to 1-h, \bh \to 1-\bh$, up to a sign
$(-1)^{2(h-\bh)}$, follows directly from \eqref{res1} and reflects the
contribution of both $\O$ and its shadow. It is straightforward to
separate $F_{h\, \bh}(\eta,{\bar \eta})$ as just the contribution 
corresponding to $\O$ and thus  \eqref{block} is just the conformal
block for a conformal primary operator labelled by $h,\bh$. A special 
case of \eqref{block} was actually given in \cite{Ferrara}, which gave
the first detailed discussion of what are now termed conformal blocks.
If $h_{12}=h_{34}=\bh_{12}=\bh_{34}=0$ $F_{0\,0}= 1$ representing the
identity.

The result \eqref{block} is in fact the leading, essentially trivial, term in 
the highly non trivial Virasoro conformal blocks, which contains contributions form
all descendants generated by $L_{-n}$ for $n=1,2,\dots$, as the 
central charge $c\to \infty$ \cite{block}. The Virasoro conformal
blocks are of course expressible as a sum over contributions 
of the form \eqref{block} for all conformal primary descendants.

Conformal invariance for the four point function for
$\vphi_1 \vphi_2 \vphi_3 \vphi_4 $ dictates
\begin{align}
& \big \langle \vphi_1(z_1,\bz_1) \,\vphi_2(z_2,\bz_2)\, \vphi_3(z_3,\bz_3) \,
\vphi_4(z_4,\bz_4)\big \rangle \nn \\
&{} = \frac{1}{z_{12}{\!}^{h_1+h_2}\,  z_{34}{\!}^{h_3+h_4}} \,
\bigg ( \frac{z_{24}}{z_{14}} \bigg )^{\! h_{12}}
\bigg ( \frac{z_{14}}{z_{13}} \bigg )^{\! h_{34}}\!\!\!
 \frac{1}{\bz_{12}{\!}^{\bh_1+\bh_2} \,\bz_{34}{\!}^{\bh_3+\bh_4}}
\bigg ( \frac{\bz_{24}}{\bz_{14}} \bigg )^{\! \bh_{12}}
\bigg ( \frac{\bz_{14}}{\bz_{13}} \bigg )^{\! \bh_{34}}\, 
\F_{1234}(\eta,{\bar \eta}) \, ,
\label{fourp}
\end{align}
and then the operator product expansion gives
\be
\F_{1234}(\eta,{\bar \eta})= 
\sum_{h,\bh} a_{h\, \bh} \, F_{h\, \bh}(\eta,{\bar \eta}) \, .
\label{exp}
\ee
The sum in \eqref{exp} includes for any $\O$ labelled by $h,\bh$
also its conjugate ${\bar \O}$ with contribution proportional to
$ F_{\bh\, h}(\eta,{\bar \eta})$,
with in general an independent  coefficient. For $\vphi_i$ real scalars, 
$h_i = \bh_i$ and $ F_{\bh\, h}(\eta,{\bar \eta})= 
F_{h\, \bh}({\bar \eta},\eta)$ and $a_{\bh\, h}= a_{h\, \bh}$ so that
\eqref{fourp} and \eqref{exp} can then be combined in the form
\begin{align}
& \big \langle \vphi_1(z_1,\bz_1) \,\vphi_2(z_2,\bz_2)\, \vphi_3(z_3,\bz_3) \,
\vphi_4(z_4,\bz_4)\big \rangle \nn \\
&{} = \frac{1}{|z_{12}|^{2(h_1+h_2)}\, |z_{34}|^{2(h_3+h_4)}} \,
\bigg ( \frac{|z_{24}|}{|z_{14}|} \bigg )^{\! 2 h_{12}}
\bigg ( \frac{|z_{14}|}{|z_{13}|} \bigg )^{\! 2 h_{34}}\, 
\sum_{h\ge \bh} a_{h\, \bh} \, \big ( F_{h\, \bh}(\eta,{\bar \eta})
+ F_{h\, \bh}({\bar \eta},\eta) \big ) \, .
\label{exp2}
\end{align}
This result is then in accord with the
form of two dimensional conformal blocks obtained in \cite{DO} where
symmetry under $\eta \leftrightarrow {\bar \eta}$ played an essential role. 
Conformal blocks in two dimensions which were odd under
$\eta \leftrightarrow {\bar \eta}$ were considered in \cite{Heem}.

As a trivial application of \eqref{exp} we may consider chiral fields
$\vphi,{\bar \vphi}$ such that
\be
 \big \langle \vphi(z_1) \,\vphi(z_2)\big \rangle = \frac{1}{z_{12}{\!}^{2h}}
\, , \quad 
\big \langle {\bar \vphi}(\bz_1) \, {\bar \vphi}(\bz_2) \big \rangle
= \frac{1}{\bz_{12}{\!}^{2\bh}} \, , \quad 2h,2\bh \in {\mathbb N} \, .
\label{twop}
\ee
The four point fuctions involving $\vphi,\vphi$ and ${\bar \vphi},{\bar \vphi}$ have
only disconnected contributions determined by \eqref{twop}. They can be cast in the
form \eqref{fourp} with
\be
\F_{\vphi\vphi{\bar \vphi}{\bar \vphi}}(\eta,{\bar \eta}) = 1 \, , \qquad
\F_{\vphi{\bar \vphi}\vphi{\bar \vphi}}(\eta,{\bar \eta}) = \eta^h \, {\bar \eta}^\bh \, .
\ee
In the first case only the identity operator contributes in the operator
product expansion, in the second since, with $h_{12} \to h, \bh_{34} \to - \bh$, 
$F_{h\, \bh}(\eta,{\bar \eta}) = \eta^h \, {\bar \eta}^\bh$, corresponding
to the conformal primary $\vphi{\bar \vphi}$.

It is interesting and less trivial to consider the role of the energy moment tensor
$T(z)$, for which $h=2,\bh=0$, and its conjugate ${\bar T}(\bz)$. In two
dimensions if $\vphi_i$ are Virasoro primary
the four point correlation functions involving $T(z)$, and also ${\bar T}(\bz)$, with 
$\vphi_1 \, \vphi_2 \, \vphi_3$ are fully determined by Ward 
identities \cite{Bel}. Hence
\begin{align}
\big \langle & T(z) \, \vphi_1(z_1,\bz_1) \,\vphi_2(z_2,\bz_2)\, 
\vphi_3(z_3,\bz_3) \, \big \rangle \nn \\
&{} = \sum_{i=1}^3 \bigg ( \frac{h_i}{(z-z_i)^2} + \frac{1}{z-z_i} \,
\frac{\pr}{\pr z_i} \bigg ) \
\big \langle  \vphi_1(z_1,\bz_1) \,\vphi_2(z_2,\bz_2)\, \vphi_3(z_3,\bz_3) \,
\big \rangle \, ,
\label{Tphi}
\end{align}
where, with the definition \eqref{threep}, we may take
\be
\big \langle  \vphi_1(z_1,\bz_1) \,\vphi_2(z_2,\bz_2)\, \vphi_3(z_3,\bz_3) \,
\big \rangle = C_{123} \, \F_{12}^{h_3 \, \bh_3}( z_3 , \bz_3) \, .
\ee
The result from \eqref{Tphi} can then be cast in the form \eqref{fourp},
with an appropriate relabelling, where 
\begin{align}
\F_{T123}(\eta,{\bar \eta}) = {}& C_{123} \, f(\eta) {\bar f}({\bar \eta})\, , 
\qquad {\bar f}({\bar \eta}) = {\bar \eta}^{\bh_1} 
( 1- {\bar \eta})^{-\bh_1 -\bh_{23} }\, ,  \nn \\
f(\eta) = {}& \eta^{h_1} ( 1-\eta)^{-h_1-h_{23}} \big ( h_1\, (1-\eta)+
h_3 \, \eta - h_2 \, \eta(1-\eta) \big ) \, .
\end{align}
To obtain an expansion as in \eqref{exp}, where in \eqref{block} we now
take $h_{12} \to 2 - h_1,  \bh_{12} \to - \bh_1$ and $ h_{34} \to h_{23} , 
\bh_{34} \to \bh_{23}$,  then since
\be
{\bar \eta}^{\bh_1} ( 1- {\bar \eta})^{-\bh_1 -\bh_{23} }
= {\bar \eta}^{\bh_1} F ( 2 \bh_1 , \bh_1 + \bh_{23} ; 2\bh_1 ; 
{\bar \eta} ) \, ,
\label{fbar}
\ee
it is sufficient to consider just
\be
f(\eta) = \sum_{n\ge 0} a_n \, \eta^{h_1+n} 
F ( 2h_1 - 2 + n , h_1+h_{23} + n ; 2h_1 + 2n ; \eta ) \, .
\label{fexp}
\ee
For $n=0,1,2$
\be
a_0 = h_1 \, , \quad a_1 = 0 \, , \quad
a_2 = \frac{h_1(h_1-1) - 3\, h_{23}{\!}^2 + (h_2+h_3)(2h_1+1)}{2(2h_1+1)} \, .
\label{a2}
\ee
The result for $n=0$, reflecting the contribution of $\vphi_1$ to the operator
product expansion of $T\vphi_1$, is determined by Ward identities. 
If $h_{23}=0$, $a_n$=0 for all odd $n$  and there is the general result
\be
a_{2p} = \bigg ( h_1 + h_2 \frac{2p}{h_1-1}(2h_1-1+2p) \bigg ) \,
\frac{(h_1-1)_{2p}}{(2h_1 - 2 +2p)_{2p}} \, .
\ee

The results \eqref{fbar} and \eqref{fexp} are a consequence of the fact that 
the conformal primary operators $\vphi_{1,n}$ present in the operator product 
expansion of $T(z) \, \vphi_1(z_1,\bz_1)$ have, for $\vphi_1$ a Virasoro primary,  
just $h=h_1+n , \bh = \bh_1$ with $n=0,2,3,\dots$. The expansion
has the form\footnote{More generally the contribution of a conformal
primary operator $\O$ to the operator product of $\vphi_1 \vphi_2$ 
is determined \cite{DO} by \eqref{ppO} and \eqref{threep} to have the form
$$
 \vphi_1(z_1,\bz_1)  \vphi_2(z_2,\bz_2) = 
 \frac{C_{12\O}}{z_{12}{\!}^{h_1+h_2}\, \bz_{12}{\!}^{\bh_1+\bh_2}}\,
z_{12}{}^h {\!}_1F_1(h+h_{12};2h;z_{12}\pr_{z_2}) 
\, \bz_{12}{}^\bh {\!}_1F_1(\bh+\bh_{12};2\bh;\bz_{12}\pr_{\bz_2}) \,
\O(z_2, \bz_2)
$$} 
\be
T(z) \, \vphi_1(0,0) = \sum_{n=0,n\ne 1}^\infty z^{n-2} 
{}_1F_1(n+2;2h_1+2n;z\,\pr) \, \vphi_{1,n}(0,0) \, ,
\label{Tp}
\ee
for $\vphi_{1,0}= h_1 \vphi_1$, and where 
$\pr^r \vphi_{1,n}(0,0)\equiv \pr_y{}^r 
\vphi_{1,n}(y,0) \big |_{y=0}$. Since $T(z) = \sum_n z^{-n-2} L_n$ 
\eqref{Tp} gives
\be
L_{-n} \vphi_1 = \sum_{r=0,r\ne n-1}^n \binom{n+1}{r} 
\frac{1}{(2h_1+2n-2r)_r} \, \pr^r \vphi_{1,n-r}\, , \ \ n=0,1,2,\dots \, ,
\ee
which can be inverted for $n=2,3,\dots$
\begin{align}
\vphi_{1,n} ={}& \sum_{r=0}^{n-2} 
\binom{n+1}{r} \frac{1}{(2h_1+2n-r-1)_r} \, (-\pr)^r L_{-n+r} \vphi_1
- \frac{(n^2-1)(2h_1+n)}{2(2h_1+n-1)_n} (-\pr)^n \vphi_1 \, .
\end{align}
As is well known if $\vphi_{1,n}$ is a Virasoro primary, so that
$L_2 \vphi_{1,n} = 0$, there are further conditions relating $h_1$
and the Virasoro central charge $c$.

Minimal models are characterised by vanishing of $\vphi_{1,n}$ for
some $n$, if  $\vphi_{1,2}=0$  then in \eqref{a2} $a_2=0$. For
$h_1=h_2$ this requires $h_3=\tfrac{1}{3}(8h_1+1)$ which is satisfied
by the Ising  model which contains two Virasoro primaries
$\sigma$, $h_\sigma=\bh_\sigma = \tfrac{1}{16}$ and $\epsilon$,
$h_\epsilon=\bh_\epsilon = \tfrac{1}{2}$ and $c=\half$. In minimal 
models correlation functions are then determined in terms of linear
partial differential equations \cite{Bel, Mattis}, for four point
functions these become ordinary differential equations in $\eta$.
Such minimal models may serve as a testing ground for applications 
of conformal bootstrap methods in higher dimensions.

For any two dimensional conformal field theory the correlations functions
of the energy momentum are universal depending only on $c$. For the
four point function
\begin{align}
& \big \langle T(z_1) \, T(z_2) \, T(z_3) \, T(z_4) \big \rangle = 
\frac{1}{z_{12}{\!}^4 \, z_{34}{\!}^4} \, \F_{TTTT}(\eta) \, , \nn \\
& \F_{TTTT}(\eta)  = \quar c^2 \big ( 1 + \eta^4 + \eta^4(1-\eta)^{-4}\big )
+ 2 c \, \eta^2 ( 1-\eta)^{-2} ( 1-\eta + \eta^2) \, .
\end{align}
The conformal partial wave expansion then takes the form\footnote{The results
can  be obtained from
$$
\eta^h = {\ts \sum_{p=0}^\infty} (-1)^p \tfrac{(h)_p{\!}^2}{p!\,(2h+p-1)_p} \, \eta^{h+p}
F(h+p,h+p;2h+2p;\eta) \, .
$$}
\begin{align}
& \F_{TTTT}(\eta)  =   \quar c^2  + \sum_{p=0}^\infty a_{2p} \, \eta^{2p+2}
F(2p+2,2p+2;4p+4;\eta) \, , \nn \\
& a_{2p} = \Big (  \tfrac{1}{144} \,  c^2 (2p-1)_6 + 2c
\big ( 1+2p(2p+3) \big ) \Big ) \frac{(2p)!\, (2p+1)!}{(4p+1)!} \, .
\end{align}

The conditions under which conformal primaries are absent from
the operator product expansion of $T(z) \vphi(0,0)$ can be obtained
from the four point function $\langle T \, \vphi \, T \, \vphi \rangle$
for which
\begin{align}
\F_{T\vphi T\vphi}(\eta)  = {}&  \half c \, 
\eta^{h+2} + \frac{\eta^h}{(1-\eta)^2}
\big ( h^2 - 2h \, \eta(1-\eta) \big ) \nn \\
={}& \sum_{n=0}^\infty C_{n} \, \eta^{h+n} F(2h+n-2,n+2;2h+2n;\eta) \, .
\end{align}
The expansion coefficients are given by
\begin{align}
C_n = {}& \Big ( \tfrac{1}{12} (n-1)_3 \, (2h)_{n+1} \, c + 
2h \big ( n(n+2h-1)+1 \big ) (2h-2)_n \Big ) \, (-1)^n 
\frac{(2h-2)_n}{(2h-2)_{2n+1}}  \nn \\
&{}+ h \big ( (n+1)(2h-2+n)-2 \big ) (n+1)! \, 
\frac{(2h-2)_n}{(2h-2)_{2n+1}}  \, .
\end{align}
The condition $C_n=0$ for suitable $c,h$ implies the absence of
the conformal primary descendant $\vphi_n$.

In higher dimensions the four point function for the energy momentum
tensor depends on the dynamical details of the particular theory
except for $\N=4$ superconformal theories at strong coupling and also 
large $N$ when it can be calculated for $d=4$ by using pure gravity
on $AdS_5$. Although this would have significant interest present
calculations \cite{Aruty} suffice only to determine the three point function.
Results based on the $AdS_4/CFT_3$ correspondence \cite{Raju} give quite
simple expressions for the four point function in three dimensions.
The four point function for the energy momentum tensor is also the natural
object to analyse in discussions of the existence of higher dimensional
conformal field theories using bootstrap methods. To achieve this 
expressions for conformal blocks with external spin 2 are essential.
The connections between conformal partial waves in even dimensions for 
external scalars suggest the possibility that this might be extended to 
non trivial external spins.
If feasible the present results in two dimensions give a very simple
starting point. 

\bigskip
\leftline{\bf Acknowledgements}
\medskip

I would  like to thank
Jo\~ao Penedones and Slava Rychkov for helpful remarks and also to Daniel
Harlow for informing me of the extensive relevant literature partially
listed in \cite{block}.

\end{document}